\documentclass{osa-article}

\journal{osajournal}

\articletype{Research Article}

\begin{document}

\title{Dissipative Kerr soliton microcombs for FEC-free optical communications over 100 channels}

\author{Shun Fujii,\authormark{1,2,5}
	 Shuya Tanaka,\authormark{1}
	 Tamiki Ohtsuka,\authormark{1}
	 Soma Kogure,\authormark{1}
	 Koshiro Wada,\authormark{1}
	 Hajime Kumazaki,\authormark{1}
	 Shun Tasaka,\authormark{1}
	 Yosuke Hashimoto,\authormark{3}
	 Yuta Kobayashi,\authormark{3}
	 Tomohiro Araki,\authormark{3}
	 Kentaro Furusawa,\authormark{4}
	 Norihiko Sekine,\authormark{4}
	 Satoki Kawanishi,\authormark{1}
	 and Takasumi Tanabe\authormark{1,*}}

\address{\authormark{1}Department of Electronics and Electrical Engineering, Faculty of Science and Technology, \\Keio University, 3-14-1, Hiyoshi, Kohoku-ku, Yokohama 223-8522, Japan\\
\authormark{2}Quantum Optoelectronics Research Team, RIKEN Center for Advanced Photonics, 2-1, Hirosawa, Wako, Saitama 351-0198, Japan\\
\authormark{3}Research Unit I, Research and Development Directorate, Japan Aerospace Exploration Agency, 2-1-1 Sengen, Tsukuba, Ibaraki 305-8505, Japan\\
\authormark{4}Terahertz Technology Research Center, National Institute of Information and Communications Technology (NICT), 4-2-1, Nukui-Kitamachi, Koganei, Tokyo 184-8795, Japan\\
\authormark{5}shun.fujii@riken.jp}

\email{\authormark{*}takasumi@elec.keio.ac.jp} 



\begin{abstract}
The demand for high-speed and highly efficient optical communication techniques has been rapidly growing due to the ever-increasing volume of data traffic. As well as the digital coherent communication used for core and metro networks, intensity modulation and direct detection (IM-DD) are still promising schemes in intra/inter data centers thanks to their low latency, high reliability, and good cost performance. In this work, we study a microresonator-based frequency comb as a potential light source for future IM-DD optical systems where applications may include replacing individual stabilized lasers with a continuous laser driven microresonator. Regarding comb line powers and spectral intervals, we compare a modulation instability comb and a soliton microcomb and provide a quantitative analysis with regard to telecom applications. Our experimental demonstration achieved a forward error correction (FEC) free operation of bit-error rate (BER) $<10^{-9}$ with a 1.45~Tbps capacity using a total of 145 lines over the entire C-band and revealed the possibility of soliton microcomb-based ultra-dense wavelength division multiplexing (WDM) with a simple, cost-effective IM-DD scheme, with a view to future practical use in data centers.
\end{abstract}

\section{Introduction}
Microresonator-based frequency combs offer a variety of applications including distance ranging \cite{Suh884,Trocha887}, low-noise microwave generation~\cite{Liang2015:high,Suh:18}, and as optical frequency synthesizers~\cite{Spencer2018}. Among numerous potential applications, microresonator frequency combs are promising for optical communication applications as multi-wavelength light sources owing to their precisely defined frequency spacing, which will potentially allow them to substitute for conventional massive laser arrays~\cite{Pfeifle2014,Marin-Palomo2017,Corcoran2020,Fulop2018}. In recent years, the capacity of coherent communication has been boosted by the adoption of highly efficient digital modulation formats such as quadrature phase-shift keying (QPSK) and quadrature amplitude modulation (QAM)~\cite{1589027}, combined with other optimization approaches including polarization-division~\cite{Chen2017} and space-division multiplexing~\cite{Richardson2013,Hu2018}. Meanwhile, microcombs have been expected to serve as spectrally efficient light sources for dense wavelength-division multiplexing (DWDM) for coherent communications~\cite{Kikuchi:16,57798,8620214,HuOxenlwe+2021+1367+1385}, specifically in long-haul transmission or metro-area networks.  A few years ago, Marin-Palomo {\it et al.} demonstrated coherent communication by using two frequency combs from chip-integrated resonators as a transmitter and a local oscillator and then realized an unprecedented data rate of up to 55~Tbps with interleaved 50~GHz carrier spacings~\cite{Marin-Palomo2017}. The development of microcomb-based high-capacity transmission is still under way and is exploiting new types of microcomb such as those employing a dark pulse~\cite{Fulop2018,Fulop:17} and a soliton crystal~\cite{Corcoran2020}, featuring a higher conversion efficiency than a bright soliton comb~\cite{doi:10.1002/lpor.201600276}. In addition to the advances in multiplexing and high-order modulation encoding, a segmentation approach has recently been demonstrated that could potentially increase the aggregate data rates~\cite{Helgason:19,9405395}. 

Although coherent communication appears to be the main microcomb-based transmission application, IM-DD is a simple and cost-effective transmission method and is still essential for short-distance communications, for example, in local area networks and intra/inter data center communications~\cite{8768325,Cheng:18,Pang:20,8918098}. We would like to emphasize that the compact microcomb-based system faces strong competition as regards such short-distance communications rather than for long-distance communications.  Moreover, when we employ IM-DD, reliable and low-latency communication is possible compared with digital coherent formats. Recently adopted 5G and 6G mobile network technologies need to support ultra-reliable and low-latency communications (URLLC), which requires a very low latency of 1~ms for packet transmission~\cite{8826541,alves2021beyond}.  To support URLLC, the wired network requires an even lower latency, which is currently a challenging task if we use digital coherent communication that needs complex post signal processing.  Therefore, there is a demand for a massive transmission capability with a simple modulation format, such as IM-DD.  Proof-of-concept experiments on IM-DD transmission were performed in the early stages of microcomb research~\cite{6218758,Wang:12}, and more practical research has recently been continuously reported~\cite{9417208,Salgals:21}.  Yet, to date there has been no comprehensive study on IM-DD transmission using low-noise soliton microcomb.

This paper studies the properties of a microcomb as a multi-wavelength light source for WDM communication applications, mainly focusing on the simple and cost-effective IM-DD method using two different microresonator platforms. We demonstrate that an error-free operation of the total 145~lines with 10~Gbit/s IM-DD encoding is possible even in extremely dense Kerr soliton microcomb generated from a 10~GHz-FSR magnesium fluoride microresonator. Here, no FEC is postulated to exploit the simplicity of the IM-DD method.  In addition, we compare the BER performance for different comb states to investigate the feasibility of using a microcomb in the IM-DD application. The other motivation behind this work is to use a theoretical analysis to fairly evaluate soliton microcomb properties for different platforms in terms of available comb power and comb bandwidth. 

The paper is organized as follows. In Section~2, we introduce the basic properties of microresonator-based frequency combs and describe the generation of an ITU Telecommunication Standardization Sector (ITU-T) aligned microcomb. In Section~3, versatile analyses are provided as a guide to the requirements of optical communications regarding soliton power and bandwidth for two microcomb platforms. In Section~4, we compare the transmission properties in different microcomb states using a silicon nitride microresonator. In Section~5, we describe ultra-dense IM-DD transmission in a 10 GHz-FSR crystalline microresonator.

\section{Frequency comb alignment for DWDM frequency grid}
\begin{figure}[h!]
	\centering\includegraphics[]{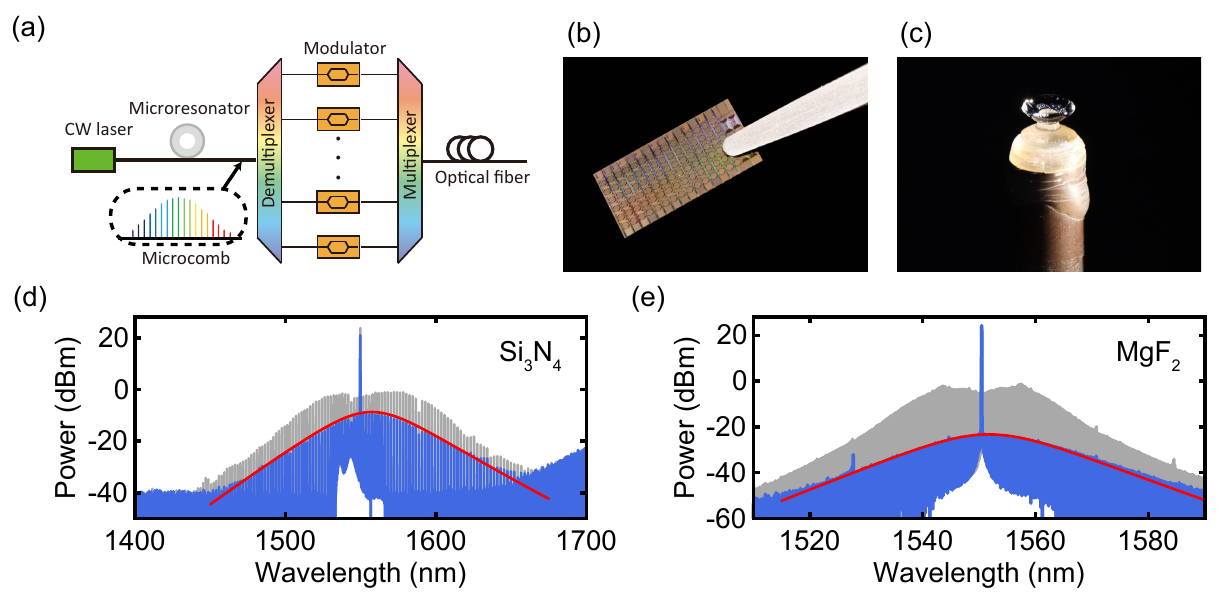}
	\caption{(a) Conceptual illustration of microcomb-based WDM communication. Each comb line works as an optical carrier for data transmission. (b) Photograph of chip-integrated silicon nitride microrings. (c) Side view of millimeter-sized magnesium fluoride microresonator. (d, e) Soliton spectrum (blue) and MI comb spectrum (gray) for $\mathrm{Si_3N_4}$ and $\mathrm{MgF_2}$ resonators, respectively. The $\mathrm{sech^2}$ fitted envelopes are denoted by red lines. The average comb power decreased significantly after the soliton formation compared with the MI comb.}
	\label{concept}
\end{figure}
\subsection{Microresonator frequency comb generation}
We first illustrate the concept of a WDM communication system in which a microresonator frequency comb is exploited as a multi-wavelength light source [Fig.~\ref{concept}(a)]. A miniature microresonator converts a continuous-wave (c.w.) laser emission into equidistant comb lines, potentially replacing a massive laser diode array with only a single-frequency pump laser. Here, we adopt two microcomb platforms; one is silicon nitride ($\mathrm{Si_3N_4}$) microrings, and the other is magnesium fluoride ($\mathrm{MgF_2}$) crystalline microresonators. Figures~\ref{concept}(b) and \ref{concept}(c) show $\mathrm{Si_3N_4}$ and $\mathrm{MgF_2}$ microresonators, respectively. Several differences, such as typical quality factors, mode volumes, and resonator sizes, determine the fundamental properties of microcombs.

Crystalline microresonators generally exhibit an extremely high Q of up to $10^9$, which allows us to obtain a narrow free-spectral range (FSR) microcomb reaching sub-10 GHz. In contrast, the usual FSR range with a $\mathrm{Si_3N_4}$ microring corresponds to 100--1000~GHz. The resonator fabrication method also makes a significant difference in that silicon nitride microresonators made from CMOS-compatible material can be massively integrated into a single chip~\cite{Levy2010}. Crystalline resonators are usually fabricated with diamond turning and subsequent polishing methods, which limit precise geometry control~\cite{GRUDININ200633}; however, recent studies have shown the possibility of fully computer-controlled crystalline resonator fabrication that can overcome this drawback~\cite{Fujii:19,Fujii:20,HAYAMA2022234}.  For crystalline microresonators, a tapered fiber coupler provides an extremely low insertion loss, otherwise a prism coupler is compatible with a system that is more robust as regards external perturbation.  Moreover, dedicated effort has led to an improvement in the Q-factors and insertion losses of chip-integrated resonators, and recently low-repetition-rate microcomb have been reported~\cite{Liu2020,Jin2021}.

Microcombs are excited by tuning a c.w. pump laser frequency to a high-Q resonance~\cite{del2007optical,Kippenberg2011microresonator}. Since the soliton state exists on the lower branch of the Kerr-bistability curve, it is generally only accessible with decreasing pump frequency~\cite{herr2014temporal,yi2015soliton}. Except for the recent discoveries of special conditions (e.g., soliton crystals~\cite{Cole2017,Karpov2019}, breathing solitons~\cite{Yu2017}, dark pulses in a normal dispersion~\cite{xue2015mode,Jang:16,8440025}), microcombs can be mainly divided into three different states, a primary comb (i.e., Turing pattern or Turing roll), chaotic modulation instability (MI), and a dissipative Kerr soliton (DKS)~\cite{PhysRevA.89.063814}. The essential difference between the former two states and a soliton state is effective pump detuning, and effective red detuning results in a low soliton conversion efficiency compared with that obtained with the MI comb~\cite{doi:10.1002/lpor.201600276}.  Figures~\ref{concept}(d) and ~\ref{concept}(e) show the observed MI comb and DKS comb in $\mathrm{Si_3N_4}$ and $\mathrm{MgF_2}$, respectively. These results reveal that the maximum comb power of the soliton state is significantly lower than that of the MI comb state. We note that the power scale (i.e., dBm) in this paper is carefully calibrated to display the actual (available) output power by considering the ratio of the optical couplers that are used to split or attenuate the optical power in the experiment. For $\mathrm{Si_3N_4}$ ring measurement, the insertion losses at a chip-facet coupling of $\sim$3 dB for each side are not considered; thus, the comb power in the bus waveguide could be almost twice the detected output power. 

\begin{figure}[t!]
	\centering\includegraphics[]{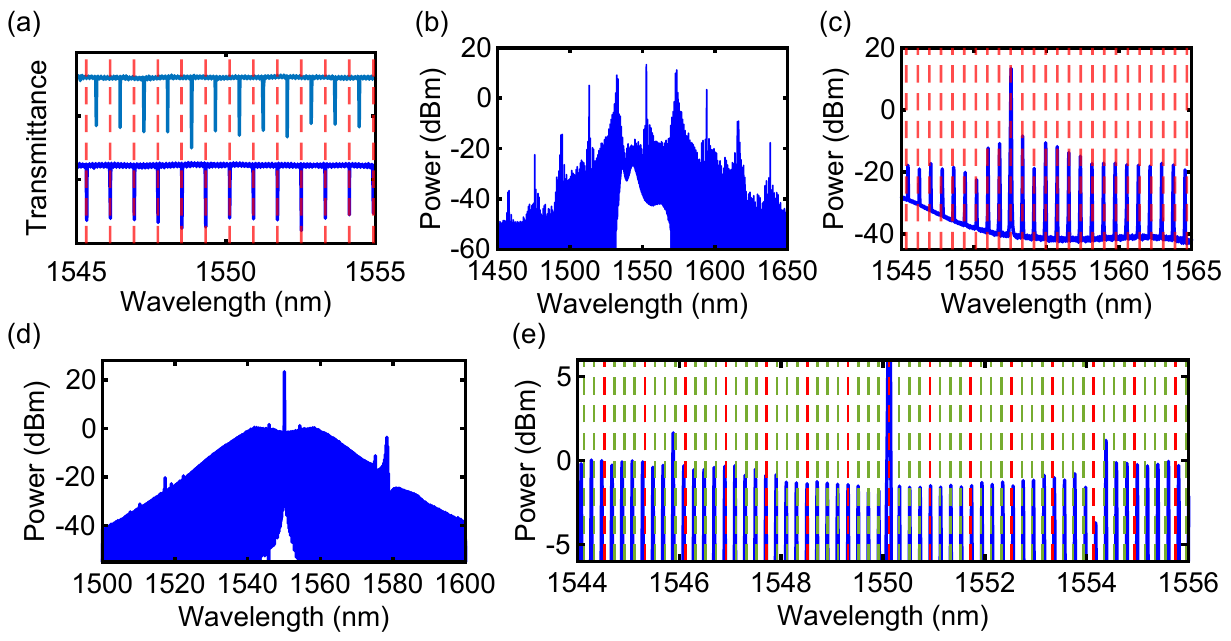}
	\caption{(a) Normalized transmittance of a 100 GHz-FSR $\mathrm{Si_3N_4}$ microresonator. The cavity resonances at room temperature (light blue) are aligned with the ITU-T frequency grid (dashed red line) via the thermo-optic effect with an efficiency of 20.2~pm/K. (b,~c) Optical spectrum of 100~GHz grid aligned $\mathrm{Si_3N_4}$ microcomb and a magnified view. (d, e) Optical spectrum of 25~GHz ITU-T grid aligned comb from an $\mathrm{MgF_2}$ microresonator and a magnified view. The dashed red and green lines represent 100~GHz and 25~GHz ITU-T frequency grids, respectively.
	}
	\label{ITU-T}
\end{figure}
\subsection{100 GHz ITU-T grid alignment in a chip-integrated microresonator}
For practical WDM applications, the frequency comb spacing should be precisely aligned with the standard spectral grid defined by the ITU-T. Since the comb spacing is essentially determined by the resonator FSR for a single-FSR microcomb, the design and fabrication of microresonators contribute fundamentally to wavelength alignment. In addition, post-fabrication resonance tuning via the thermo-optic effect is a useful method for realizing precise alignment on the DWDM grid.

To demonstrate a 100~GHz grid-aligned microcomb, we use a $\mathrm{Si_3N_4}$ microresonator with a radius of 228~\textmu m, and the cross-sectional dimensions are 0.8~\textmu m (height) $\times$ 1.6 \textmu m (width). The measured FSR of the cavity is 99.8~GHz, and the absolute wavelength offset between a cold resonance and the ITU-T grid is approximately 0.5~nm at room temperature (RT). We tune the resonances by controlling a thermo-electric cooler (TEC) mounted under a cavity sample to bridge such large wavelength differences. As a result, a resonant wavelength shift of 20.2~pm/K is achieved, which agrees with the previously reported value~\cite{Xue:16}. By carefully adjusting the temperature, we aligned the resonances with the ITU-T grid within an error of 0.1~nm over 20~channels as shown in Fig.~\ref{ITU-T}(a). Finally, the grid-aligned microcomb is also demonstrated with a pump power of 500~mW [Figs.~\ref{ITU-T}(b) and \ref{ITU-T}(c)].

\subsection{25 GHz ITU-T grid alignment in a crystalline microresonator}
Resonator fabrication based on a lithographic technique facilitates precise FSR adjustment. However, precise control will present a technical challenge, particularly for high-Q crystalline microresonators because millimeter-sized crystalline resonators are generally fabricated by a manual grinding and polishing method. Therefore, during fabrication considerable effort is needed to obtain the desired resonator diameter. Here, we use an $\mathrm{MgF_2}$ resonator with an FSR of 25.24~GHz. The generated comb spectrum is shown in Figs.~\ref{ITU-T}(d) and \ref{ITU-T}(e). Significantly, we obtain an output power of $\sim$0~dBm for each spectrum line throughout almost the entire C-band (1535 to 1565~nm), and 20~comb lines are adjusted with the ITU-T grid with less than a 10\% frequency mismatch.

\section{Comb line power and bandwidth analysis for telecom applications}

In this section, we begin with a discussion of comb power and comb bandwidth in different microcomb platforms for telecom applications. Although there are several resonator platforms that can be used for microcomb generation, a straightforward approach is to choose from a typical free-spectral range (comb mode-spacing), comb line power, and comb bandwidth. Since the comb line power and bandwidth of a soliton microcomb can be analytically derived from the Lugiato-Lefever equation (LLE)~\cite{lugiato1987spatial}, it is helpful to quantitatively compare typical soliton comb properties on two different platforms. A detailed description and corresponding equations are provided in Appendix.~A.

Figure~\ref{colormap} shows the contour maps for the maximum comb line power and a 3-dB spectral bandwidth in silicon nitride [Figs.~\ref{colormap}(a)-\ref{colormap}(c)] and magnesium fluoride microresonators [Figs.~\ref{colormap}(d)-\ref{colormap}(f)]. We chose typical values for each resonator as a parameter space. It should also be noted that slight over-coupling is assumed (coupling coefficient $\eta$=0.7), and an effective mode area is fixed at 1.5~\textmu m$^2$ and 125~\textmu m$^2$ for $\mathrm{Si_3N_4}$ and $\mathrm{MgF_2}$ resonators, respectively. Figures~\ref{colormap}(a) and \ref{colormap}(d) indicate that resonators with a large FSR and low Q yield higher comb line powers, whereas the minimum pump power needed to maintain a soliton, i.e. the minimum soliton existence power, increases greatly due to their low-Q properties. Hence, the pump power and loaded Q directly determine the available comb mode spacings. 

Then, we changed the group velocity dispersion (GVD) $\beta_2$ with a fixed FSR and a pulse duration of 100~GHz and 41~fs for $\mathrm{Si_3N_4}$ and 25~GHz and 200~fs for $\mathrm{MgF_2}$ resonator [Figs.~\ref{colormap}(b) and \ref{colormap}(e)]. Although a maximum comb line power is essentially independent of a pulse duration, the pulse width is tentatively fixed so that the pump detuning parameter can be assigned to the horizontal axis of the map. This in turn helps us deduce the position in the map directly from the experimental parameter. We still recognize that a strong anomalous dispersion can enhance the comb line power in return for requiring a higher pump power. The blue dashed line represents the soliton existence limit, which corresponds to the minimum pump detuning derived from the bistability condition. As a result, a strong anomalous dispersion relaxes the soliton existence range and makes it possible to realize higher comb line powers, given that sufficient pump power is provided. Nevertheless, it is important to note that the 3~dB comb bandwidth decreases under a strong anomalous dispersion as shown in Figs.~\ref{colormap}(c) and \ref{colormap}(f). As a countermeasure, large pump detuning compensates for the comb bandwidth since the higher pumping power extends the maximum detuning, which is determined by the cavity linewidth and threshold power for parametric oscillation~\cite{yi2015soliton}. Other conditions (e.g., mode crossings, thermal effect, and pump tuning speed) influence the accessible maximum detuning (soliton step length) in practice, and thus careful adjustment is essential to facilitate the realization of a broadband soliton microcomb~\cite{herr2014temporal}. This trade-off relation plays a key role in deciding the comb power and bandwidth, both of which are fundamental requirements for optical communication applications.

\begin{figure}[th!]
	\centering\includegraphics[]{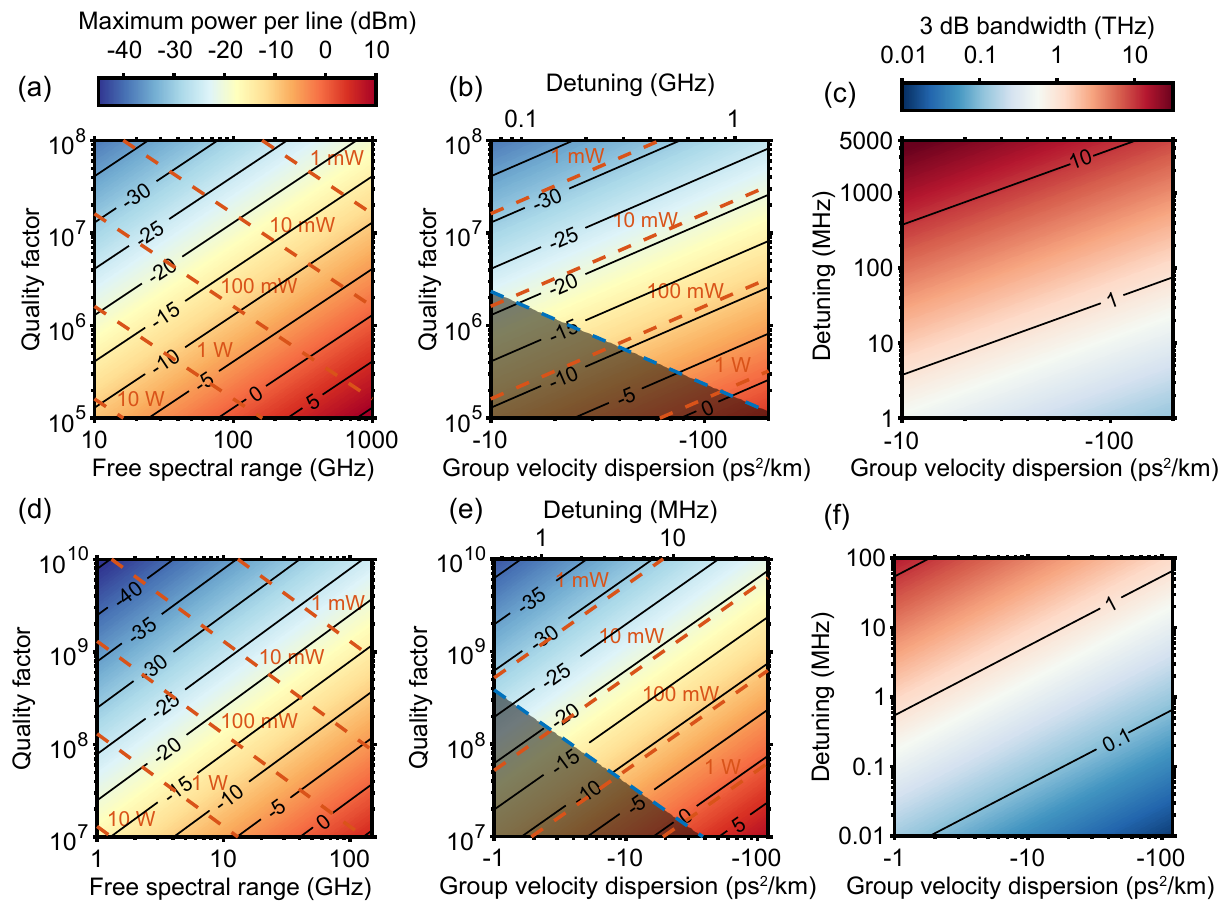}
	\caption{Contour map for maximum power per comb line and 3 dB bandwidth of soliton comb. The dashed red lines indicate the minimum pump power needed for the existence of soliton, and the dashed blue lines represent the existence limit given by the bistability condition (the shaded gray region corresponds to an area where no soliton exists). The pump wavelength of 1550 nm and coupling $\eta=0.7$ are common in the calculation. Other nonlinear effects (e.g., Raman induced soliton self-frequency shift) and higher-order dispersion are neglected. (a-c) Silicon nitride ($\mathrm{Si_3N_4}$) microresonator. A soliton pulse duration of 41 fs and GVD $\beta_2$ of $-100~\mathrm{ps^2/km}$ are assumed for (a), and a 41~fs soliton pulse and FSR of 100~GHz are assumed for (b). (d-f) Magnesium fluoride ($\mathrm{MgF_2}$) microresonator. A soliton pulse duration of 200~fs and GVD $\beta_2$ of $-10~\mathrm{ps^2/km}$ are assumed for (d), and a 200~fs soliton pulse and an FSR of 25~GHz are assumed for (e). The 3 dB bandwidth yields the pulse duration from the time-bandwidth product (TBP) of 0.315 for a $\mathrm{sech^2}$-shaped soliton pulse.}
	\label{colormap}
\end{figure}

\section{Comparison of BER properties in different microcomb states}
\begin{figure}[t!]
	\centering\includegraphics[]{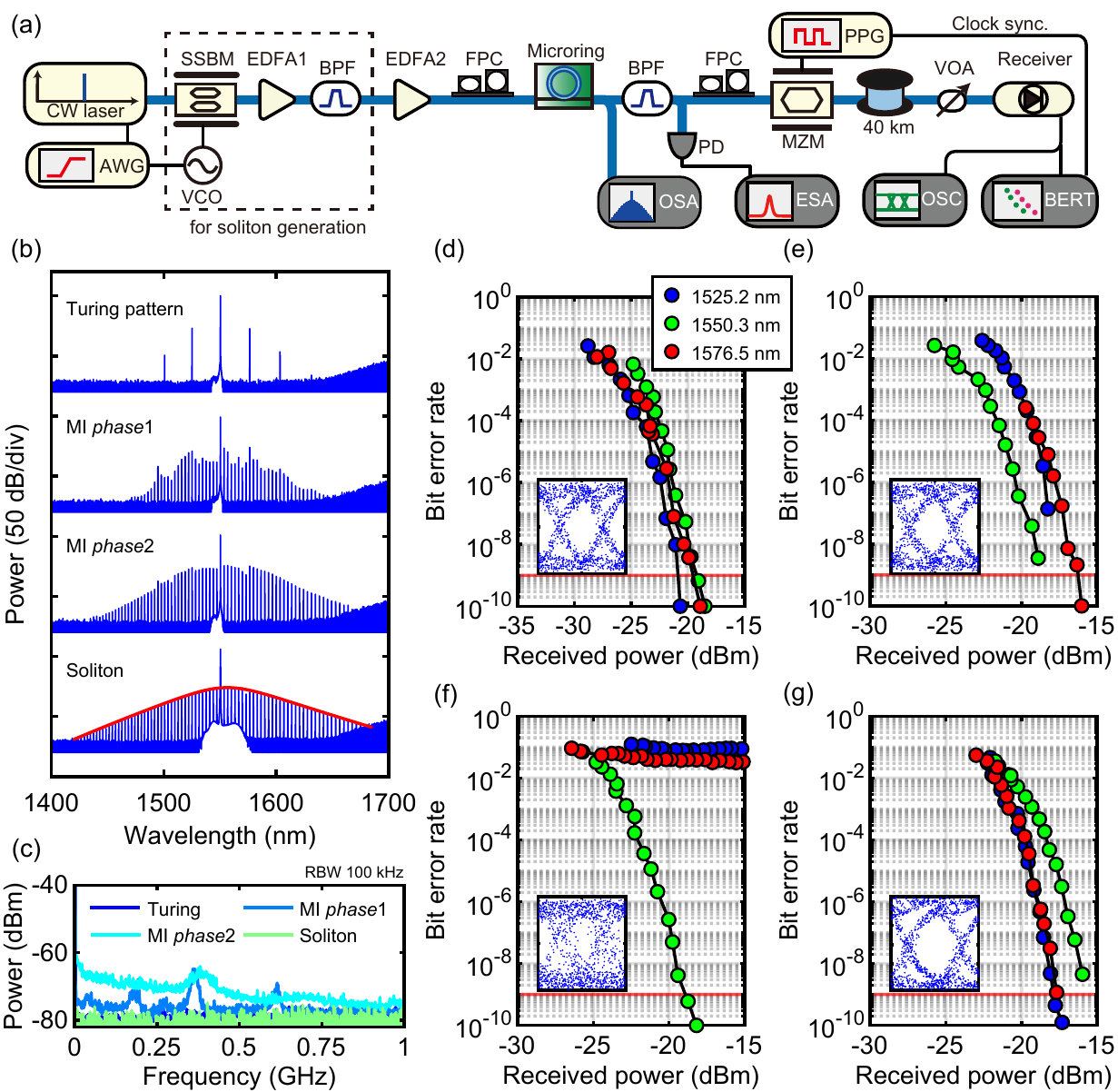}
	\caption{(a) Setup for a microcomb transmission experiment. The components inside the dashed box are used for soliton generation. AWG: Arbitrary waveform generator, SSBM: Single-sideband modulator, VCO: Voltage-controlled oscillator, BPF: Band-pass filter, EDFA: Erbium-doped fiber amplifier, FPC: Fiber polarization controller, PD: Photodetector, PPG: Pulse-pattern generator, MZM: Mach-Zehnder modulator, VOA: Variable optical attenuator, OSA: Optical spectrum analyzer, ESA: Electrical spectrum analyzer, OSC: Oscilloscope, BERT: Bit error rate tester. (b) Optical spectra of microcomb in different states. The resonator exhibits an FSR of 400~GHz and a Q-factor of $1.3\times10^6$. (c) RF noise spectra of each comb state. (d)-(g) Bit error rate versus received power for each comb state: (d) Turing pattern, (e) MI {\it phase1}, (f) MI {\it phase2}, (g) Soliton comb. The three comb lines including a pump light were selected for BER measurement. Insets show eye pattern diagrams at a wavelength of 1525.2~nm (The horizontal scale is 200~ps). The red lines correspond to a BER of $10^{-9}$, often considered the minimum acceptable BER for telecom applications.}
	\label{SiN}
\end{figure}

This section compares the BER performance in different comb states (i.e., Turing pattern, MI comb, and soliton comb) to investigate which states are suitable for IM-DD transmission. In our experiment, different microcomb states are generated using a  $\mathrm{Si_3N_4}$ resonator with an FSR of 400~GHz. Figure~\ref{SiN}(a) shows an experimental setup for microcomb transmission. Turing and MI combs are easily accessible by slowly tuning the pump laser wavelength (Santec TSL-710), but for soliton generation, the fast-scanning method via a single-sideband modulator is employed to circumvent the strong thermal effect~\cite{PhysRevLett.121.063902}. The optical spectra of different states are shown in Fig.~\ref{SiN}(b) when the off-chip pump power is 1~W (0.5~W on chip). Here, MI combs are classified into two different states by the amount of pump detuning; we refer to the initially observed MI comb as ``{\it phase1}'' and the second one as ``{\it phase2}''. In practice, the Turing pattern comb gradually changes to {\it phase1} by reducing the pump-cavity detuning, and {\it phase2} emerges only after {\it phase1} by further reducing the detuning.

For a transmission experiment, selected comb lines are modulated at 10 Gbit/s with non-return-to-zero (NRZ) coding. As shown in Fig.~\ref{SiN}(a), the comb line is filtered out via a band-pass filter before an intensity modulator with careful control of the input polarization. The modulated carriers are transmitted through a 40-km fiber spool and received by a transceiver connected to a digital sampling oscilloscope (Agilent 86100A with 86103A module) and a bit error rate tester (Anritsu 1762C). Additionally, an electric spectrum analyzer (Advantest R3273) is used to monitor the RF beat noise of the generated comb line [Fig.~\ref{SiN}(c)]. In the following experiment, three selected lines (1525.2~nm, 1550.3~nm, and 1576.5~nm) are chosen as transmitted carriers, which correspond to the pump and primary sidebands of a Turing pattern comb, for comparison.

The results for BER as a function of received power and the corresponding eye diagrams are shown in Figs.~\ref{SiN}(d)-\ref{SiN}(g). We found that a Turing pattern comb exhibits excellent BER properties, indicating an error-free operation of $10^{-9}$ ($10^{-10}$ is our measurement limit). These results are in accordance with the fact that primary combs exhibit robust low-noise properties compared with MI combs, which are more suitable for optical communication systems. Indeed, earlier work has reported a high-capacity coherent data transmission of up to 144 Gbit/s per channel using a Turing pattern~\cite{PhysRevLett.114.093902}. Next, we perform the same measurement on the MI comb {\it phase1} and {\it phase2} to investigate the degree to which the noise of chaotic MI combs affects the BER performance. As shown in Figs.~\ref{SiN}(d) and \ref{SiN}(e), the BER performance of {\it phase1} degraded compared with that of the Turing pattern comb despite the comb power levels being the same. Furthermore, the performance of {\it phase2} deteriorated significantly in terms of both BER and eye diagram [Fig.~\ref{SiN}(f)]. We confirm that error-free transmission was not achieved in the MI comb {\it phase2} although the comb spectrum is almost flat top and is denser than the preceding two states. Intuitively, the non-negligible intensity noise of an MI comb is considered to limit the BER performance, as shown in Fig.~\ref{SiN}(c). Multiple peaks in the RF spectrum originate from sub-comb formation and subsequent beating between combs with different mode spacings. Similar results have been observed in earlier experiments~\cite{herr2012universal,Wang:12,Pfeifle2014}.

Finally, we perform the measurement on a soliton comb. DKS combs are mode-locked and feature coherent and low-noise properties; therefore, recent transmission experiments have been focusing on utilizing soliton states. Nevertheless, the low pump-comb conversion efficiency is an unavoidable drawback of the soliton comb, which results in it having a low optical signal-to-noise ratio (OSNR) as a light source. Indeed, the power ratio of the pump to comb line is approximately 25.4~dB in an MI comb {\it phase2} but can reach 32.7 dB in a soliton comb, as shown in Fig.~\ref{SiN}(a). Figure~\ref{SiN}(g) confirms that the BER performance improved due to its low-noise property in spite of the relatively low comb power. We also measured the BER for other comb lines between 1525~nm and 1550~nm in addition to the three lines we report, and all the comb lines exhibited error-free operation. 
 
 The important findings are the Turing pattern and soliton comb can provide low-noise light sources for transmission applications, whereas the smaller the pump-cavity detuning changes, the worse the comb noise property becomes in chaotic MI combs. Some previous demonstrations of microcomb-based optical transmission are considered to represent operation in a relatively low-noise comb state, such as an MI comb {\it phase1}~\cite{6218758,Wang:12,Pfeifle2014}.

\begin{figure}[t!]
	\centering\includegraphics[]{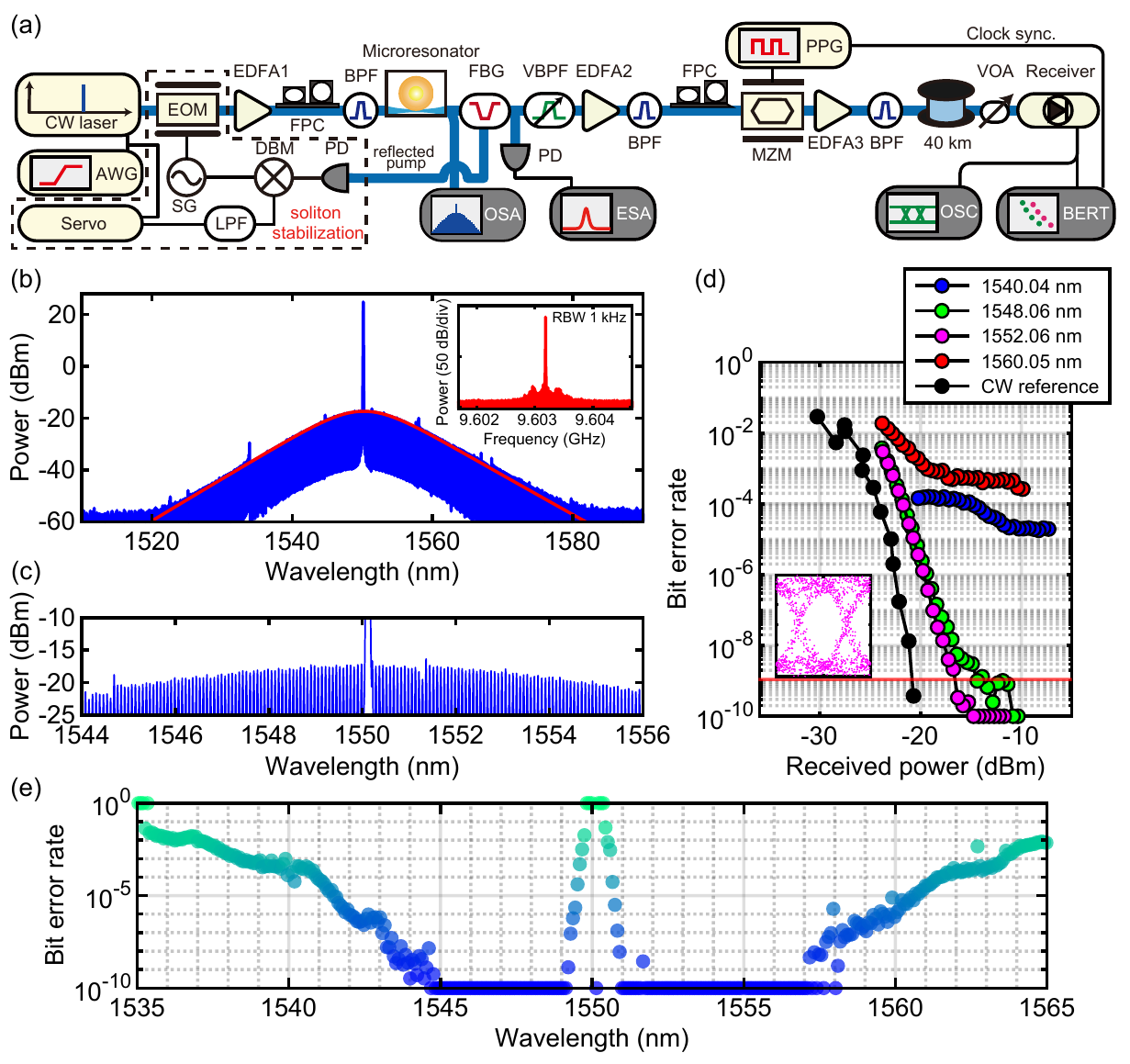}
	\caption{(a) Experimental setup for soliton microcomb transmission. The components inside the dashed box are used for soliton stabilization via the Pound-Drever-Hall (PDH) method. SG: Signal generator, DBM: Double-balanced mixer, LPF: Low-pass filter, FBG: Fiber Bragg grating, VBPF: Variable band-pass filter. (b) Optical spectrum of a single soliton state with a $\mathrm{sech^2}$-envelope (red line). The inset shows the RF spectrum of the soliton repetition frequency (9.60~GHz). The loaded Q-factor of the pump mode was $2.4\times10^8$. (c) Magnified view of a soliton comb spectrum. (d) Bit error rate versus received power for four selected comb lines and a CW reference at 10 Gbit/s. The two lines (1548.06 nm and 1552.06 nm) show error-free operation, whereas the BERs of other comb lines are clumped around $10^{-3}$--$10^{-5}$ due to the weak line power of the native soliton comb. The inset shows an eye pattern diagram at a wavelength of 1552.06~nm (The horizontal scale is 200~ps). (e) Bit error rate spectrum of a total of 386 lines covering the C-band, of which 145 lines exhibit error-free operation. The low BER lines in the vicinity of the pump are attributed to ASE noise caused by an EDFA (EDFA1).
	}
	\label{MgF2}
\end{figure}

\section{Soliton microcomb-based ultra-dense IM-DD transmission}
\subsection{Experimental result}
Although a narrow comb mode spacing boosts the transmission capacity of DWDM communication, the line powers obtained from such a large-size resonator are much lower than those obtained with a wider FSR comb, as shown in Fig.~\ref{colormap}. MI combs are superior to soliton combs in terms of comb line power, but on the other hand, the strong intensity noise leads to the critical degradation of transmission properties. This section highlights ultra-dense soliton microcombs as optical carriers and investigates the transmission properties of each comb line over the entire C-band.

A 7-mm diameter $\mathrm{MgF_2}$ microresonator yields a single soliton with an FSR of $~$10 GHz, which ultimately enhances the transmission capacity with an NRZ encoding with data encoding rate of 10~Gbit/s. The experimental setup is shown in Fig.~\ref{MgF2}(a). A pump laser is amplified by an EDFA and $\sim$350~mW optical power is coupled via an optical tapered fiber. A single soliton is generated via laser frequency tuning with an adequate scan speed, and the soliton state is stabilized by locking effective pump detuning with a servo controller, resulting in stable soliton operation for several hours. In contrast to the experiment using a $\mathrm{Si_3N_4}$ microresonator, a 0.1~kHz-linewidth single-frequency laser is used as a pump laser, and an optical tapered fiber is employed as a coupler. Figures~\ref{MgF2}(b) and \ref{MgF2}(c) show the observed soliton comb spectra. The $\mathrm{sech}^2$ fitting to the obtained soliton spectrum yields a 3~dB bandwidth of 1.22~THz, corresponding to a pulse duration of 259~fs. The comb lines within the C-band total 386 lines, and the maximum comb line power through an output coupler is $-$17~dBm. The comb FSR of 9.6~GHz corresponding to the soliton repetition rate was characterized using a high-bandwidth photoreceiver and an electrical spectrum analyzer [inset in~Fig.~\ref{SiN}(b)]. 

For the transmission demonstration, we modulate the observed soliton comb with 10 Gbit/s NRZ coding and monitor the BERs after a 40-km fiber transmission. Figure~\ref{MgF2}(d) shows the BERs for the selected channels. The two comb lines (1548.06~nm and 1552.06~nm) exhibit error-free operation, whereas the BERs of the other comb lines remain at $10^{-3}$--$10^{-5}$ due to their weak line powers, resulting in a low OSNR. The BERs for all the comb channels covering the entire C-band are shown in Fig.~\ref{MgF2}(e), indicating that a total of 145 carriers achieved error-free operation, i.e., BER $=10^{-9}$. This result suggests the realization of an ultrahigh transmission capacity of up to 1.45~Tbps by employing a DWDM configuration even with the IM-DD method. The spectral efficiency yields 1~bit/s/Hz without assuming the use of FEC techniques, which could alleviate the comb line power requirement. To the best of our knowledge, our result represents the densest carrier spacing in microcomb-based optical communication and the first demonstration of a simple, low-cost IM-DD data transmission experiment using a soliton microcomb. 

\subsection{Discussion}
While the maximum modulation rate is limited by a comb spacing of $\sim$10 GHz to avoid the crosstalk between the other channels, advanced modulation formats (e.g., PAM4, and PAM8), could increase the total transmission capacity as demonstrated by a very recent study of data transmission with wavelength selective switching using integrated soliton microcombs~\cite{Raja2021}.  Yet, encoding more bits per symbol requires a higher OSNR for each comb line~\cite{Marin-Palomo:20}. The achievable OSNR can be determined by the comb line powers, namely the source-limited, thus demanding a high-power microcomb. Besides, line-to-line power variations induce the channel-dependent OSNR, which can be improved by shaping the comb spectrum. Dispersion engineering such that resonators possess strong anomalous dispersion will be a crucial technique for obtaining high power per line, even though it sacrifices the soliton comb bandwidth. In addition, the achievable dispersion is ultimately restricted by the material and resonator geometry~\cite{FujiiTanabe+2020+1087+1104}; for instance, an anomalous dispersion of $-200~\mathrm{ps^2/km}$ is the maximum attainable value for currently used silicon nitride rings~\cite{Helgason:19}. Therefore, it is still a challenge to obtain a tens of gigahertz mode-spacing, broad-bandwidth, and high-power microcomb. Another possibility is to use a soliton crystal~\cite{Corcoran2020}, a dark pulse~\cite{Fulop2018} or a laser cavity-soliton~\cite{Bao2019}. A mode-locked dark pulse provides a higher conversion efficiency than a soliton comb, and specifically, a disordered soliton crystal exhibits intrinsic high efficiency and power flatness around the pump mode~\cite{Cole2017}.

\section{Conclusion}
In conclusion, we investigated the properties of a microcomb as a candidate for a multi-wavelength light source with the IM-DD method. Starting with the introduction of microcombs on two different platforms, we discussed the advantages and disadvantages in terms of scalability, coupling method, and frequency spacing. A multifaceted analysis based on LLE provides a comprehensive strategy for deciding a platform from the viewpoint of line power and comb bandwidth, which govern the fundamental performance of transmission applications. While a microcomb exhibits distinct noise-level performance depending on the pump and resonator condition, we characterized each comb state through BER measurements using a chip-integrated microring resonator. Then, we demonstrated extremely dense data transmission using a 10 GHz mode spacing soliton microcomb in a crystalline microresonator. The results anticipate a net rate of 1.45 Tbit/s without FEC, even though a narrow FSR comb exhibits relatively low line power compared with a comb with a wider mode spacing. The results encourage the use of the IM-DD configuration rather than coherent communication, particularly in intra/inter data center networks where ultralow latency and cost-effectiveness are still of primary importance.

\clearpage

\section*{Appendix}
\appendix
\section{Mathematical description of analytical soliton expression}
Theoretical analyses of soliton power and bandwidth have been conducted in several previous studies~\cite{Matsko:13,yi2015soliton,herr2014temporal}. Here, we briefly describe the derivation of the comb line power and bandwidth by following the literature, which yields the expressions for plotting Fig.~\ref{colormap}. The soliton average power and pulse width are given by
\begin{equation}\label{psolc3}
	P_\mathrm{sol} = \frac{2\eta A_\mathrm{eff}}{n_2Q}\sqrt{-2n_0c\beta_2\cdot\delta\omega},
\end{equation}
\begin{equation}\label{solduration}
	\tau = \sqrt{-\frac{c\beta_2}{2n_0\cdot\delta\omega}},
\end{equation}
where $\eta$ is the coupling ratio, $A_\mathrm{eff}$ is the effective mode area, $n_2$ is the nonlinear index, $n_0$ is the refractive index, $c$ is the speed of light in a vacuum, and $\delta \omega=(\omega_0-\omega_p)$ is the pump detuning. Eliminating the detuning term from Eq.~(\ref{psolc3}) and Eq.~(\ref{solduration}) gives the following expression, 
\begin{equation}\label{psoltau}
	P_\mathrm{sol} = -\frac{2c\eta A_\mathrm{eff}}{n_2\tau}\frac{\beta_2}{Q}.
\end{equation}
The optical spectrum of a soliton and each comb line power is given by assuming a hyperbolic secant form as follows,
\begin{equation}\label{combpower}
	P(\Delta\omega) = -\frac{\pi c}{2}\frac{\eta  A_\mathrm{eff}}{n_2}\frac{\beta_2D_1}{Q} \mathrm{sech}^2\left(\frac{\pi\tau}{2}\Delta\omega\right),
\end{equation}
where $D_1/2\pi$ is the resonator FSR, $\Delta\omega$ is the relative frequency from the pump, and thus $P(\Delta \omega=0)$ corresponds to the maximum comb line power. The soliton emerges only when satisfying the bistability condition, which defines the minimum pump detuning as,
\begin{equation}\label{deomin}
	\delta\omega_\mathrm{min} = 2\pi\frac{\sqrt{3}}{2}\Delta\nu,
\end{equation}
where $\Delta\nu$ corresponds to the cavity linewidth in hertz. Using the Q-factor, this relation is rewritten as,
\begin{equation}
	 Q = \frac{\sqrt{3}\nu_0}{2(\delta\omega_\mathrm{min}/2\pi)}.
\end{equation}
Meanwhile, the maximum detuning for the existence of a soliton is given by,
\begin{equation}\label{deomaxxx}
	\delta\omega_\mathrm{max} = 2\pi\frac{\pi^2P_\mathrm{in}}{16P_\mathrm{th}}\Delta\nu.
\end{equation}
Thus, regarding the cavity linewidth, the detuning range for the existence of a soliton is given by
\begin{equation}
	\frac{\sqrt{3}}{2}\Delta\nu < \frac{\delta\omega}{2\pi} < \frac{\pi^2P_\mathrm{in}}{16P_\mathrm{th}}\Delta\nu,
\end{equation}
where  $P_\mathrm{in}$ is the input power and $P_\mathrm{th}$ is the threshold power for parametric oscillation,
\begin{equation}\label{thresh}
	P_\mathrm{th} = \frac{\pi n_0\omega_0A_\mathrm{eff}}{4\eta n_2}\frac{1}{D_1Q^2}.
\end{equation}
The maximum detuning can be rewritten by substituting Eq.~(\ref{thresh}) for Eq.~(\ref{deomaxxx}) as
\begin{equation}
	\frac{\delta\omega_\mathrm{max}}{2\pi} = \frac{\eta n_2P_\mathrm{in}}{8n_0A_\mathrm{eff}}D_{1}Q,
\end{equation}
which suggests that the maximum detuning is proportional to the cavity FSR and Q-factor. At the maximum detuning, furthermore, eliminating the detuning term from Eq.~(\ref{solduration}) yields the minimum pump power needed for a soliton to exist,
\begin{equation}\label{pmin}
 P_\mathrm{in}^\mathrm{min} = -\frac{2cA_\mathrm{eff}}{\pi\eta n_2}\frac{\beta_2}{QD_1}\frac{1}{\tau^2}.
\end{equation}

\begin{backmatter}
	\bmsection{Funding}
	
	\bmsection{Acknowledgments}
	 S. Fujii is supported by the RIKEN Special Postdoctoral Program. The authors thank Dr. H. Tsuda for technical support.
	\bmsection{Disclosures}
	The authors declare no conflicts of interest.
	\bmsection{Data Availability Statement}
	Data underlying the results presented in this paper are not publicly available at this time but may be obtained from the authors upon reasonable request.

\end{backmatter}


\begin{thebibliography}{10}
	\newcommand{\enquote}[1]{``#1''}
	
	\bibitem{Suh884}
	M.-G. Suh and K.~J. Vahala, \enquote{Soliton microcomb range measurement,}
	{\protect\JournalTitle{Science}} \textbf{359}, 884--887 (2018).
	
	\bibitem{Trocha887}
	P.~Trocha, M.~Karpov, D.~Ganin, M.~H.~P. Pfeiffer, A.~Kordts, S.~Wolf,
	J.~Krockenberger, P.~Marin-Palomo, C.~Weimann, S.~Randel, W.~Freude, T.~J.
	Kippenberg, and C.~Koos, \enquote{Ultrafast optical ranging using
		microresonator soliton frequency combs,} {\protect\JournalTitle{Science}}
	\textbf{359}, 887--891 (2018).
	
	\bibitem{Liang2015:high}
	W.~Liang, D.~Eliyahu, V.~S. Ilchenko, A.~A. Savchenkov, A.~B. Matsko,
	D.~Seidel, and L.~Maleki, \enquote{High spectral purity {K}err frequency comb
		radio frequency photonic oscillator,} {\protect\JournalTitle{Nat. Commun.}}
	\textbf{6}, 7957 (2015).
	
	\bibitem{Suh:18}
	M.-G. Suh and K.~Vahala, \enquote{Gigahertz-repetition-rate soliton
		microcombs,} {\protect\JournalTitle{Optica}} \textbf{5}, 65--66 (2018).
	
	\bibitem{Spencer2018}
	D.~T. Spencer, T.~Drake, T.~C. Briles, J.~Stone, L.~C. Sinclair, C.~Fredrick,
	Q.~Li, D.~Westly, B.~R. Ilic, A.~Bluestone, N.~Volet, T.~Komljenovic,
	L.~Chang, S.~H. Lee, D.~Y. Oh, M.-G. Suh, K.~Y. Yang, M.~H.~P. Pfeiffer,
	T.~J. Kippenberg, E.~Norberg, L.~Theogarajan, K.~Vahala, N.~R. Newbury,
	K.~Srinivasan, J.~E. Bowers, S.~A. Diddams, and S.~B. Papp, \enquote{An
		optical-frequency synthesizer using integrated photonics,}
	{\protect\JournalTitle{Nature}} \textbf{557}, 81--85 (2018).
	
	\bibitem{Pfeifle2014}
	J.~Pfeifle, V.~Brasch, M.~Lauermann, Y.~Yu, D.~Wegner, T.~Herr, K.~Hartinger,
	P.~Schindler, J.~Li, D.~Hillerkuss, R.~Schmogrow, C.~Weimann, R.~Holzwarth,
	W.~Freude, J.~Leuthold, T.~J. Kippenberg, and C.~Koos, \enquote{Coherent
		terabit communications with microresonator {K}err frequency combs,}
	{\protect\JournalTitle{Nature Photonics}} \textbf{8}, 375--380 (2014).
	
	\bibitem{Marin-Palomo2017}
	P.~Marin-Palomo, J.~N. Kemal, M.~Karpov, A.~Kordts, J.~Pfeifle, M.~H.~P.
	Pfeiffer, P.~Trocha, S.~Wolf, V.~Brasch, M.~H. Anderson, R.~Rosenberger,
	K.~Vijayan, W.~Freude, T.~J. Kippenberg, and C.~Koos,
	\enquote{Microresonator-based solitons for massively parallel coherent
		optical communications,} {\protect\JournalTitle{Nature}} \textbf{546},
	274--279 (2017).
	
	\bibitem{Corcoran2020}
	B.~Corcoran, M.~Tan, X.~Xu, A.~Boes, J.~Wu, T.~G. Nguyen, S.~T. Chu, B.~E.
	Little, R.~Morandotti, A.~Mitchell, and D.~J. Moss, \enquote{Ultra-dense
		optical data transmission over standard fibre with a single chip source,}
	{\protect\JournalTitle{Nature Communications}} \textbf{11}, 2568 (2020).
	
	\bibitem{Fulop2018}
	A.~F{\"u}l{\"o}p, M.~Mazur, A.~Lorences-Riesgo, {\'O}.~B. Helgason, P.-H. Wang,
	Y.~Xuan, D.~E. Leaird, M.~Qi, P.~A. Andrekson, A.~M. Weiner, and
	V.~Torres-Company, \enquote{High-order coherent communications using
		mode-locked dark-pulse {K}err combs from microresonators,}
	{\protect\JournalTitle{Nature Communications}} \textbf{9}, 1598 (2018).
	
	\bibitem{1589027}
	D.-S. Ly-Gagnon, S.~Tsukamoto, K.~Katoh, and K.~Kikuchi, \enquote{Coherent
		detection of optical quadrature phase-shift keying signals with carrier phase
		estimation,} {\protect\JournalTitle{Journal of Lightwave Technology}}
	\textbf{24}, 12--21 (2006).
	
	\bibitem{Chen2017}
	Z.-Y. Chen, L.-S. Yan, Y.~Pan, L.~Jiang, A.-L. Yi, W.~Pan, and B.~Luo,
	\enquote{Use of polarization freedom beyond polarization-division
		multiplexing to support high-speed and spectral-efficient data transmission,}
	{\protect\JournalTitle{Light: Science {\&} Applications}} \textbf{6},
	e16207--e16207 (2017).
	
	\bibitem{Richardson2013}
	D.~J. Richardson, J.~M. Fini, and L.~E. Nelson, \enquote{Space-division
		multiplexing in optical fibres,} {\protect\JournalTitle{Nature Photonics}}
	\textbf{7}, 354--362 (2013).
	
	\bibitem{Hu2018}
	H.~Hu, F.~Da~Ros, M.~Pu, F.~Ye, K.~Ingerslev, E.~Porto~da Silva,
	M.~Nooruzzaman, Y.~Amma, Y.~Sasaki, T.~Mizuno, Y.~Miyamoto, L.~Ottaviano,
	E.~Semenova, P.~Guan, D.~Zibar, M.~Galili, K.~Yvind, T.~Morioka, and L.~K.
	Oxenl{\o}we, \enquote{Single-source chip-based frequency comb enabling
		extreme parallel data transmission,} {\protect\JournalTitle{Nature
			Photonics}} \textbf{12}, 469--473 (2018).
	
	\bibitem{Kikuchi:16}
	K.~Kikuchi, \enquote{Fundamentals of coherent optical fiber communications,}
	{\protect\JournalTitle{J. Lightwave Technol.}} \textbf{34}, 157--179 (2016).
	
	\bibitem{57798}
	C.~Brackett, \enquote{Dense wavelength division multiplexing networks:
		principles and applications,} {\protect\JournalTitle{IEEE Journal on Selected
			Areas in Communications}} \textbf{8}, 948--964 (1990).
	
	\bibitem{8620214}
	V.~Torres-Company, J.~Schröder, A.~Fülöp, M.~Mazur, L.~Lundberg, Ó.~B.
	Helgason, M.~Karlsson, and P.~A. Andrekson, \enquote{Laser frequency combs
		for coherent optical communications,} {\protect\JournalTitle{Journal of
			Lightwave Technology}} \textbf{37}, 1663--1670 (2019).
	
	\bibitem{HuOxenlwe+2021+1367+1385}
	H.~Hu and L.~K. Oxenløwe, \enquote{Chip-based optical frequency combs for
		high-capacity optical communications,} {\protect\JournalTitle{Nanophotonics}}
	\textbf{10}, 1367--1385 (2021).
	
	\bibitem{Fulop:17}
	A.~F\"{u}l\"{o}p, M.~Mazur, A.~Lorences-Riesgo, T.~A. Eriksson, P.-H. Wang,
	Y.~Xuan, D.~E. Leaird, M.~Qi, P.~A. Andrekson, A.~M. Weiner, and
	V.~Torres-Company, \enquote{Long-haul coherent communications using
		microresonator-based frequency combs,} {\protect\JournalTitle{Opt. Express}}
	\textbf{25}, 26678--26688 (2017).
	
	\bibitem{doi:10.1002/lpor.201600276}
	X.~Xue, P.-H. Wang, Y.~Xuan, M.~Qi, and A.~M. Weiner, \enquote{Microresonator
		{K}err frequency combs with high conversion efficiency,}
	{\protect\JournalTitle{Laser Photon. Rev.}} \textbf{11}, 1600276 (2017).
	
	\bibitem{Helgason:19}
	\'{O}skar B.~Helgason, A.~F\"{u}l\"{o}p, J.~Schr\"{o}der, P.~A. Andrekson,
	A.~M. Weiner, and V.~Torres-Company, \enquote{Superchannel engineering of
		microcombs for optical communications,} {\protect\JournalTitle{J. Opt. Soc.
			Am. B}} \textbf{36}, 2013--2022 (2019).
	
	\bibitem{9405395}
	M.~Mazur, M.-G. Suh, A.~Fülöp, J.~Schröder, V.~Torres-Company, M.~Karlsson,
	K.~Vahala, and P.~Andrekson, \enquote{High spectral efficiency coherent
		superchannel transmission with soliton microcombs,}
	{\protect\JournalTitle{Journal of Lightwave Technology}} \textbf{39},
	4367--4373 (2021).
	
	\bibitem{8768325}
	Q.~Hu, M.~Chagnon, K.~Schuh, F.~Buchali, and H.~Bülow, \enquote{{IM}/{DD}
		beyond bandwidth limitation for data center optical interconnects,}
	{\protect\JournalTitle{Journal of Lightwave Technology}} \textbf{37},
	4940--4946 (2019).
	
	\bibitem{Cheng:18}
	Q.~Cheng, M.~Bahadori, M.~Glick, S.~Rumley, and K.~Bergman, \enquote{Recent
		advances in optical technologies for data centers: a review,}
	{\protect\JournalTitle{Optica}} \textbf{5}, 1354--1370 (2018).
	
	\bibitem{Pang:20}
	X.~Pang, O.~Ozolins, R.~Lin, L.~Zhang, A.~Udalcovs, L.~Xue, R.~Schatz,
	U.~Westergren, S.~Xiao, W.~Hu, G.~Jacobsen, S.~Popov, and J.~Chen,
	\enquote{200 {G}bps/lane {IM}/{DD} technologies for short reach optical
		interconnects,} {\protect\JournalTitle{J. Lightwave Technol.}} \textbf{38},
	492--503 (2020).
	
	\bibitem{8918098}
	X.~Zhou, R.~Urata, and H.~Liu, \enquote{Beyond 1 {T}b/s intra-data center
		interconnect technology: {IM-DD} or {C}oherent?}
	{\protect\JournalTitle{Journal of Lightwave Technology}} \textbf{38},
	475--484 (2020).
	
	\bibitem{8826541}
	A.~Ghosh, A.~Maeder, M.~Baker, and D.~Chandramouli, \enquote{5{G} {E}volution:
		A view on 5{G} cellular technology beyond 3{GPP} release 15,}
	{\protect\JournalTitle{IEEE Access}} \textbf{7}, 127639--127651 (2019).
	
	\bibitem{alves2021beyond}
	H.~Alves, G.~D. Jo, J.~Shin, C.~Yeh, N.~H. Mahmood, C.~Lima, C.~Yoon,
	N.~Rahatheva, O.-S. Park, S.~Kim \emph{et~al.}, \enquote{Beyond 5{G} {URLLC}
		{E}volution: New service modes and practical considerations,}
	{\protect\JournalTitle{arXiv:2106.11825}}  (2021).
	
	\bibitem{6218758}
	J.~S. Levy, K.~Saha, Y.~Okawachi, M.~A. Foster, A.~L. Gaeta, and M.~Lipson,
	\enquote{High-performance silicon-nitride-based multiple-wavelength source,}
	{\protect\JournalTitle{IEEE Photonics Technology Letters}} \textbf{24},
	1375--1377 (2012).
	
	\bibitem{Wang:12}
	P.-H. Wang, F.~Ferdous, H.~Miao, J.~Wang, D.~E. Leaird, K.~Srinivasan, L.~Chen,
	V.~Aksyuk, and A.~M. Weiner, \enquote{Observation of correlation between
		route to formation, coherence, noise, and communication performance of {K}err
		combs,} {\protect\JournalTitle{Opt. Express}} \textbf{20}, 29284--29295
	(2012).
	
	\bibitem{9417208}
	S.~Spolitis, R.~Murnieks, L.~Skladova, T.~Salgals, A.~V. Andrianov, M.~P.
	Marisova, G.~Leuchs, E.~A. Anashkina, and V.~Bobrovs, \enquote{{IM}/{DD}
		{WDM}-{PON} communication system based on optical frequency comb generated in
		silica whispering gallery mode resonator,} {\protect\JournalTitle{IEEE
			Access}} \textbf{9}, 66335--66345 (2021).
	
	\bibitem{Salgals:21}
	T.~Salgals, J.~Alnis, R.~Murnieks, I.~Brice, J.~Porins, A.~V. Andrianov, E.~A.
	Anashkina, S.~Spolitis, and V.~Bobrovs, \enquote{Demonstration of a fiber
		optical communication system employing a silica microsphere-based {OFC}
		source,} {\protect\JournalTitle{Opt. Express}} \textbf{29}, 10903--10913
	(2021).
	
	\bibitem{Levy2010}
	J.~S. Levy, A.~Gondarenko, M.~A. Foster, A.~C. Turner-Foster, A.~L. Gaeta, and
	M.~Lipson, \enquote{{CMOS}-compatible multiple-wavelength oscillator for
		on-chip optical interconnects,} {\protect\JournalTitle{Nat. Photonics}}
	\textbf{4}, 37--40 (2010).
	
	\bibitem{GRUDININ200633}
	I.~S. Grudinin, A.~B. Matsko, A.~A. Savchenkov, D.~Strekalov, V.~S. Ilchenko,
	and L.~Maleki, \enquote{Ultra high {Q} crystalline microcavities,}
	{\protect\JournalTitle{Opt. Commun.}} \textbf{265}, 33 -- 38 (2006).
	
	\bibitem{Fujii:19}
	S.~Fujii, S.~Tanaka, M.~Fuchida, H.~Amano, Y.~Hayama, R.~Suzuki, Y.~Kakinuma,
	and T.~Tanabe, \enquote{Octave-wide phase-matched four-wave mixing in
		dispersion-engineered crystalline microresonators,}
	{\protect\JournalTitle{Opt. Lett.}} \textbf{44}, 3146--3149 (2019).
	
	\bibitem{Fujii:20}
	S.~Fujii, Y.~Hayama, K.~Imamura, H.~Kumazaki, Y.~Kakinuma, and T.~Tanabe,
	\enquote{All-precision-machining fabrication of ultrahigh-{Q} crystalline
		optical microresonators,} {\protect\JournalTitle{Optica}} \textbf{7},
	694--701 (2020).
	
	\bibitem{HAYAMA2022234}
	Y.~Hayama, S.~Fujii, T.~Tanabe, and Y.~Kakinuma, \enquote{Theoretical approach
		on the critical depth of cut of single crystal {M}g{F}$_2$ and application to
		a microcavity,} {\protect\JournalTitle{Precision Engineering}} \textbf{73},
	234--243 (2022).
	
	\bibitem{Liu2020}
	J.~Liu, E.~Lucas, A.~S. Raja, J.~He, J.~Riemensberger, R.~N. Wang, M.~Karpov,
	H.~Guo, R.~Bouchand, and T.~J. Kippenberg, \enquote{Photonic microwave
		generation in the {X}- and {K}-band using integrated soliton microcombs,}
	{\protect\JournalTitle{Nature Photonics}} \textbf{14}, 486--491 (2020).
	
	\bibitem{Jin2021}
	W.~Jin, Q.-F. Yang, L.~Chang, B.~Shen, H.~Wang, M.~A. Leal, L.~Wu, M.~Gao,
	A.~Feshali, M.~Paniccia, K.~J. Vahala, and J.~E. Bowers,
	\enquote{Hertz-linewidth semiconductor lasers using {CMOS}-ready
		ultra-high-{Q} microresonators,} {\protect\JournalTitle{Nature Photonics}}
	\textbf{15}, 346--353 (2021).
	
	\bibitem{del2007optical}
	P.~Del'Haye, A.~Schliesser, O.~Arcizet, T.~Wilken, R.~Holzwarth, and T.~J.
	Kippenberg, \enquote{Optical frequency comb generation from a monolithic
		microresonator,} {\protect\JournalTitle{Nature}} \textbf{450}, 1214--1217
	(2007).
	
	\bibitem{Kippenberg2011microresonator}
	T.~J. Kippenberg, R.~Holzwarth, and S.~A. Diddams,
	\enquote{Microresonator-based optical frequency combs,}
	{\protect\JournalTitle{Science}} \textbf{332}, 555--559 (2011).
	
	\bibitem{herr2014temporal}
	T.~Herr, V.~Brasch, J.~Jost, C.~Wang, N.~Kondratiev, M.~Gorodetsky, and T.~J.
	Kippenberg, \enquote{Temporal solitons in optical microresonators,}
	{\protect\JournalTitle{Nat. Photonics}} \textbf{8}, 145--152 (2014).
	
	\bibitem{yi2015soliton}
	X.~Yi, Q.-F. Yang, K.~Y. Yang, M.-G. Suh, and K.~J. Vahala, \enquote{Soliton
		frequency comb at microwave rates in a high-{Q} silica microresonator,}
	{\protect\JournalTitle{Optica}} \textbf{2}, 1078--1085 (2015).
	
	\bibitem{Cole2017}
	D.~C. Cole, E.~S. Lamb, P.~Del'Haye, S.~A. Diddams, and S.~B. Papp,
	\enquote{Soliton crystals in {K}err resonators,}
	{\protect\JournalTitle{Nature Photonics}} \textbf{11}, 671--676 (2017).
	
	\bibitem{Karpov2019}
	M.~Karpov, M.~H.~P. Pfeiffer, H.~Guo, W.~Weng, J.~Liu, and T.~J. Kippenberg,
	\enquote{Dynamics of soliton crystals in optical microresonators,}
	{\protect\JournalTitle{Nature Physics}} \textbf{15}, 1071--1077 (2019).
	
	\bibitem{Yu2017}
	M.~Yu, J.~K. Jang, Y.~Okawachi, A.~G. Griffith, K.~Luke, S.~A. Miller, X.~Ji,
	M.~Lipson, and A.~L. Gaeta, \enquote{Breather soliton dynamics in
		microresonators,} {\protect\JournalTitle{Nature Communications}} \textbf{8},
	14569 (2017).
	
	\bibitem{xue2015mode}
	X.~Xue, Y.~Xuan, Y.~Liu, P.-H. Wang, S.~Chen, J.~Wang, D.~E. Leaird, M.~Qi, and
	A.~M. Weiner, \enquote{Mode-locked dark pulse {K}err combs in
		normal-dispersion microresonators,} {\protect\JournalTitle{Nat. Photonics}}
	\textbf{9}, 594--600 (2015).
	
	\bibitem{Jang:16}
	J.~K. Jang, Y.~Okawachi, M.~Yu, K.~Luke, X.~Ji, M.~Lipson, and A.~L. Gaeta,
	\enquote{Dynamics of mode-coupling-induced microresonator frequency combs in
		normal dispersion,} {\protect\JournalTitle{Opt. Express}} \textbf{24},
	28794--28803 (2016).
	
	\bibitem{8440025}
	S.~{Fujii}, Y.~{Okabe}, R.~{Suzuki}, T.~{Kato}, A.~{Hori}, Y.~{Honda}, and
	T.~{Tanabe}, \enquote{Analysis of mode coupling assisted {K}err comb
		generation in normal dispersion system,} {\protect\JournalTitle{IEEE
			Photonics J.}} \textbf{10}, 1--11 (2018).
	
	\bibitem{PhysRevA.89.063814}
	C.~Godey, I.~V. Balakireva, A.~Coillet, and Y.~K. Chembo, \enquote{Stability
		analysis of the spatiotemporal {L}ugiato-{L}efever model for {K}err optical
		frequency combs in the anomalous and normal dispersion regimes,}
	{\protect\JournalTitle{Phys. Rev. A}} \textbf{89}, 063814 (2014).
	
	\bibitem{Xue:16}
	X.~Xue, Y.~Xuan, C.~Wang, P.-H. Wang, Y.~Liu, B.~Niu, D.~E. Leaird, M.~Qi, and
	A.~M. Weiner, \enquote{Thermal tuning of {K}err frequency combs in silicon
		nitride microring resonators,} {\protect\JournalTitle{Opt. Express}}
	\textbf{24}, 687--698 (2016).
	
	\bibitem{lugiato1987spatial}
	L.~A. Lugiato and R.~Lefever, \enquote{Spatial dissipative structures in
		passive optical systems,} {\protect\JournalTitle{Phys. Rev. Lett.}}
	\textbf{58}, 2209 (1987).
	
	\bibitem{PhysRevLett.121.063902}
	J.~R. Stone, T.~C. Briles, T.~E. Drake, D.~T. Spencer, D.~R. Carlson, S.~A.
	Diddams, and S.~B. Papp, \enquote{Thermal and nonlinear dissipative-soliton
		dynamics in {K}err-microresonator frequency combs,}
	{\protect\JournalTitle{Phys. Rev. Lett.}} \textbf{121}, 063902 (2018).
	
	\bibitem{PhysRevLett.114.093902}
	J.~Pfeifle, A.~Coillet, R.~Henriet, K.~Saleh, P.~Schindler, C.~Weimann,
	W.~Freude, I.~V. Balakireva, L.~Larger, C.~Koos, and Y.~K. Chembo,
	\enquote{Optimally coherent {K}err combs generated with crystalline
		whispering gallery mode resonators for ultrahigh capacity fiber
		communications,} {\protect\JournalTitle{Phys. Rev. Lett.}} \textbf{114},
	093902 (2015).
	
	\bibitem{herr2012universal}
	T.~Herr, K.~Hartinger, J.~Riemensberger, C.~Wang, E.~Gavartin, R.~Holzwarth,
	M.~Gorodetsky, and T.~J. Kippenberg, \enquote{Universal formation dynamics
		and noise of {K}err-frequency combs in microresonators,}
	{\protect\JournalTitle{Nat. Photonics}} \textbf{6}, 480--487 (2012).
	
	\bibitem{Raja2021}
	A.~S. Raja, S.~Lange, M.~Karpov, K.~Shi, X.~Fu, R.~Behrendt, D.~Cletheroe,
	A.~Lukashchuk, I.~Haller, F.~Karinou, B.~Thomsen, K.~Jozwik, J.~Liu,
	P.~Costa, T.~J. Kippenberg, and H.~Ballani, \enquote{Ultrafast optical
		circuit switching for data centers using integrated soliton microcombs,}
	{\protect\JournalTitle{Nature Communications}} \textbf{12}, 5867 (2021).
	
	\bibitem{Marin-Palomo:20}
	P.~Marin-Palomo, J.~N. Kemal, T.~J. Kippenberg, W.~Freude, S.~Randel, and
	C.~Koos, \enquote{Performance of chip-scale optical frequency comb generators
		in coherent {WDM} communications,} {\protect\JournalTitle{Opt. Express}}
	\textbf{28}, 12897--12910 (2020).
	
	\bibitem{FujiiTanabe+2020+1087+1104}
	S.~Fujii and T.~Tanabe, \enquote{Dispersion engineering and measurement of
		whispering gallery mode microresonator for {K}err frequency comb generation,}
	{\protect\JournalTitle{Nanophotonics}} \textbf{9}, 1087--1104 (2020).
	
	\bibitem{Bao2019}
	H.~Bao, A.~Cooper, M.~Rowley, L.~Di~Lauro, J.~S. Totero~Gongora, S.~T. Chu,
	B.~E. Little, G.-L. Oppo, R.~Morandotti, D.~J. Moss, B.~Wetzel, M.~Peccianti,
	and A.~Pasquazi, \enquote{Laser cavity-soliton microcombs,}
	{\protect\JournalTitle{Nature Photonics}} \textbf{13}, 384--389 (2019).
	
	\bibitem{Matsko:13}
	A.~B. Matsko and L.~Maleki, \enquote{On timing jitter of mode locked {K}err
		frequency combs,} {\protect\JournalTitle{Opt. Express}} \textbf{21},
	28862--28876 (2013).
	
\end{thebibliography}


\end{document}